\documentclass[aps,prd,amsmath,amssymb,showpacs]{revtex4}
\usepackage{graphics,epsfig,epsf,psfrag}
\usepackage{bm}
\usepackage{color}
\usepackage{amsmath}
\usepackage{amssymb}
\usepackage{amsfonts}
\usepackage{keyval,graphicx}
\usepackage{textcomp,wasysym}

\newcommand{\x}{X(3872)}

\newcommand{\intq}{\int\!\frac{\mathrm d^3q}{(2\pi)^3}}
\newcommand{\mev}{~\mathrm{MeV}}

\begin{document}
\title{$X(3872)$ in a hot pion bath}
\author{Martin Cleven$^{1}$\footnote{{\it E-mail address:} cleven@fqa.ub.edu},
        Volodymyr K. Magas$^{1}$\footnote{{\it E-mail address:} vladimir@fqa.ub.edu},
        and Angels Ramos$^{1}$\footnote{{\it E-mail address:} ramos@fqa.ub.edu}
        }
        
\affiliation{$^1$ Departament de Fisica Quantica i Astrofisica and Institut de Ciencies del Cosmos\\
        Universitat de Barcelona, 08028-Barcelona, Spain}


\begin{abstract}
Right since its discovery in 2003 by the Belle collaboration, establishing the nature of the X(3872) meson has been one of the main priorities in the field of quarkonium physics. Not qualifying as a conventional $c\bar{c}$ state, the multiquark structure of this exotic meson has received very different interpretations, ranging from a compact tetraquark configuration to an extended  $D\bar D^*+c.c.$ molecule. In this work we explore the effect that a hot pion bath may have in the properties of the $\x$, assuming this state to be a $D\bar D^*+c.c.$ molecule. We derive the finite temperature effects on the $\x$ from a coupled channels unitarized amplitude, obtained including the  properties of the charmed mesons under such conditions.
We find that the $\x$ develops a subtantial width, of the order of a few tens of MeV, in hot pionic environments at temperatures 100-150~MeV, and its nominal mass moves above the $D D^*$ threshold.  The fact that the $\x$ in a hot pion gas may no longer be a narrow resonance needs to be considered in the estimation of production yields in relativistic heavy-ion collisions. 
\end{abstract}
\maketitle

\section{Introduction}

The discovery of the $\x$ shortly after the turn of the century~\cite{Choi:2003ue},  followed by its confirmation by various collaborations \cite{Acosta:2003zx,Abazov:2004kp,Aubert:2004ns,Aaij:2011sn}, was a milestone in quarkonium physics. 
For the first time a charmonium state was seen that is at odds with the interpretation as a conventional $c \bar c$ state.
Until this day its true nature has not been determined unambigiously. 
While a conventional charmonium state is ruled out, various alternative interpretations have been given.
On one side, the vicinity of the $\x$ mass to the $D D^{*}$ threshold inspires the models which treat the $\x$ as a molecular
$D D^{*}$ bound state with a small binding energy \cite{mol1,mol2}.
Other models treat the $\x$ as a bound diquark and antidiquark (tetraquark) state \cite{tetra1,tetra2}.  These are the most known and most frequently discussed models of the $\x$, although there are also more exotic ideas, for example an approach of Ref. \cite{combi_2_4} treats the $\x$ as an admixture of two and four-quark states. 
For an overview of the situation and a detailed discussion of the various models see the reviews~\cite{Guo:2017jvc,Brambilla:2010cs,Esposito:2014rxa} and references therein.

In most cases the various interpretations study whether the generation of the $\x$ within their model fits the charmonium spectrum or whether the branching fractions for two- or three-body decays match the experimental observations. And so far it has not been possible to determine the structure of the $\x$, since the existing data can be well explained by quite different models. 
The situation, however, may change if one applies the above discussed molecular or tetraquark models  to describe the production of
exotic charmonium in hadronic reactions, for example $pp$ \cite{Bignamini:2009sk,Carvalho:2015nqf}, or in relativistic heavy ion collisions \cite{Cho:2010db,Cho:2011ew,Cho:2017dcy,Cho:2013rpa,Abreu:2016qci}. 

High energy heavy ion collisions offer an interesting scenario to study the production
of multiquark states in general, and the $\x$ resonance in particular \cite{Cho:2010db,Cho:2011ew,Cho:2017dcy,Cho:2013rpa,Abreu:2016qci} . Assuming that the production mechanism proceeds from the coalescence of its constituents, the ExHIC collaboration showed that the yield of the $\x$ strongly depends on its internal structure \cite{Cho:2010db,Cho:2011ew,Cho:2017dcy}. In particular, at RHIC and LHC energies the $\x$ production yield is about 20 times smaller for a tetraquark configuration than for a molecular structure. 
 
This difference can become even stronger if one takes into account the further $\x$ tetraquark evolution in the hadronic phase \cite{Cho:2013rpa,Abreu:2016qci}. The point is that if $\x$ is a tetraquark it will be produced (by coalescence of quarks and antiquarks) at the end of the hadronization of the quark-gluon plasma generated in the collision; while if $\x$ is a molecular state it will be formed by hadron coalescence much later at the end of hadronic phase evolution, at the so called kinetic freeze out. Thus, after being produced at the end of the quark gluon plasma phase, the $\x$ tetraquark interacts with other hadrons during the expansion of the hadronic matter. The $\x$ can be destroyed in collisions with the comoving light mesons, mostly pions, for example in the reaction $\x+\pi \rightarrow D+ \bar D$; and at the same time some $\x$ particles can be generated through the inverse reactions, i.e. $D+ \bar D \rightarrow \x+\pi$.  The detailed study of all possible hadronic reactions performed in \cite{Torres:2014fxa,Abreu:2016qci}
has shown that at highest RHIC energies the final production yield of the $\x$ considered as a tetraquark state becomes about 80 times  smaller than that for a  $D D^{*}$ molecule \cite{Abreu:2016qci}, a result which relied on the production yield of the molecular $\x$ state calculated in the hadron coalescence model of the ExHIC collaboration \cite{Cho:2010db,Cho:2011ew}.

Based on these studies, one may therefore tend to interpret the fact that the $\x$  particle has so far not been seen in heavy ion experiments in favor of its tetraquark composition. However, the hadron coalescence model used to calculate the production yield for the molecular $\x$ state \cite{Cho:2010db,Cho:2011ew,Cho:2017dcy} does not consider its possible modification in the hot hadronic (actually pionic) medium. This is the subject of this letter. 

We study the behaviour of $\x$ in a finite-temperature pion bath, which we consider to be the first level approximation of the matter generated in ultrarelativistic heavy ion collision,  under the assumption that it is a molecular state formed by charmed meson interactions. From our previous study on how charmed mesons behave under such conditions, see Ref. \cite{Cleven:2017fun}, we know that the charmed $D$ and $D^*$ mesons acquire a substantial width, reaching for example values of the order of $30-40$ MeV at a temperature $T=150$ MeV.
Since we assume that the $\x$ is generated by the interactions of these charmed mesons, the modification of their spectral functions will necessarily affect the properties of this composite state and, consequently, its production yields.

%
\section{Framework}\label{sec:framework}

In this section we describe the framework employed to obtain the properties of the $\x$ resonance, generated from the interactions of the charmed mesons in a pionic medium. These interactions are described by a combination of $SU(4)$ effective Lagrangians  introduced by Gamermann et al. in Refs.~\cite{Gamermann:2006nm,Gamermann:2007fi} and the Imaginary Time Formalism (ITF) \cite{galekapustabook,lebellac} in a self-consistent approach. Details of the formalism can be found in our earlier work \cite{Cleven:2017fun}, where the $D$ and $D^*$ properties in a hot pionic gas were obtained from the $D\pi$ and $D^*\pi$ interactions.

Consistently, in the present work we also employ the $SU(4)$ effective Lagrangians to obtain the  $D D^*/D_s D_s^*$ coupled channel interaction leading to the generation of the $\x$ with $J^{PC}=1^{++}$ and $I=0$ quantum numbers. The scattering potential between a vector and a pseudscalar meson is given by
\begin{equation}
 V_{ij}(s,t,u) = - \frac{\xi _{ij}}{4f^2}(s-u) \, \epsilon\cdot \epsilon^\prime \ ,
 \label{eq:pot}
 \end{equation}
where $s$ and $u$ are the usual Mandelstam variables and $\epsilon$, $ \epsilon^\prime$ are the polarization vectors of the vector mesons involved in the vertex. In the case studied here, where the relevant channels are $ D\bar D^*+c.c.$ and $D_s \bar D_s^*+c.c.$,  the parameter $f$, which in the SU(3) sector stands for the pion decay constant, is replaced by that of the heavy $D$-meson, $f_D=165$~MeV. The matrix of coefficients $\xi _{ij}$ is given by:
\begin{equation}
 \xi = \left( \begin{array}{cc}  -\psi-2 & -\sqrt2  \\  -\sqrt2 & -\psi-1    \end{array}\right) \ ,
\end{equation}
where $\psi$ is a SU(4)-breaking parameter, defined as $\psi=-1/3+4/3(m_L/m_{H^\prime})$, which accounts for the different mass of the mesons that can be exchanged in a t-channel diagram that would be approximated by the point-like potential of Eq.~(\ref{eq:pot}). In the present  $ D D^*/D_s D_s^* $ coupled channel case one may exchange light type mesons ($\rho,\omega,\phi,K^*$), for which we assume a common mass $m_L\sim 800$ MeV, or the heavy $J/\Psi$ one, with mass $ m_{H^\prime}\sim3000$ MeV. 

The corresponding $s$-wave projection of this potential in then used as the kernel for the Bethe-Salpeter equation, which implicitly sums the multiple meson-meson scattering processes to all orders.
Within the on-shell formalism this simplifies to a simple algebraic equation that can be easily solved as
\begin{equation}\label{Eq:T}
 T= (1 - VG)^{-1}V\,\vec \epsilon\cdot \vec \epsilon\,^\prime \ ,
\end{equation}
for the scattering of vector mesons off pseudoscalar ones.
The diagonal matrix $G$ contains the two-meson loops and reads
\begin{eqnarray}\label{Eq:G_Vacuum}
 G_{ii}(s) &=& {\rm i} \int\!\frac{\mathrm d^4q}{(2\pi)^4} \frac{1}{[q^2-m_1^2+{\rm i}\varepsilon][(P-q)^2-m_2^2+{\rm i}\varepsilon]}\\
 &= &{\frac{1}{16\pi ^2}}\biggr[ \alpha+\log{\frac {m_1^2 }{ \mu ^2}}+{\frac{m_2^2-m_1^2+s}{ 2s}}\log{\frac{m_2^2}{ m_1^2}}+\nonumber\\ 
  &&{\frac{{\rm p}}{\sqrt{s}}}\Big( \log{\frac{s-m_2^2+m_1^2+2{\rm p}\sqrt{s} }{ -s+m_2^2-m_1^2+2{\rm p}\sqrt{s}}}+\log{\frac{s+m_2^2-m_1^2+2{\rm p}\sqrt{s} }{ -s-m_2^2+m_1^2+  2{\rm p}\sqrt{s}}}\Big)\biggr] \ ,
\end{eqnarray} 
where the index $i$ refers to the pair of mesons with masses $m_1$ and $m_2$, ${\rm p}$ is the on-shell three-momentum of the mesons in the c.m. frame
and $P^2=s$. The scale $\mu$ is set to 1.5 GeV and the subtraction constant used here is $\alpha=-1.26$. With this model and parameters the $\x$ emerges as a pole of the scattering amplitude a couple of MeV below the averaged $D  D^*$ threshold.

The properties of the $\x$ resonance in a hot pionic gas will be derived from a temperature dependent amplitude obtained by solving Eq.~(\ref{Eq:T}) with a two-meson loop function $G$ that incorporates the medium effects. Within ITF, the meson-meson loop at finite temperature reads
\begin{eqnarray}\label{Eq:G_TVac}
 G_{MM^\prime}(P^0,\vec P;T) = \intq \int\!\mathrm d\omega\int\!\mathrm d\omega^\prime \frac{S_M(\omega,\vec q;T)S_{M^\prime}(\omega^\prime,\vec P - \vec q;T)}{P^0-\omega-\omega^\prime+{\rm i}\varepsilon}     [1+f(\omega,T)+f(\omega^\prime,T)] \ ,
\end{eqnarray}
where $f(\omega,T)=[{\rm exp}(\omega/T)-1]^{-1}$ is the meson Bose distribution function at temperature $T$, while $S_M(\omega,\vec q;T)$ denotes the spectral function of meson $M$,
\begin{equation}
 S_M(\omega,\vec q;T) = -(1/\pi)\mathrm{Im}(D_M(\omega,\vec q;T))\ ,
\end{equation}
which is related to the imaginary part of the meson propagator given by
 \begin{equation}
 D_M(\omega,\vec q;T) = [\omega^2-\vec q\,^2-m_M^2-\Pi_M(\omega,\vec q;T)]^{-1} \ .
 \label{eq:prop}
\end{equation}
The quantity $\Pi_M(\omega,\vec q;T)$ is the meson self-energy, which is obtained from closing the pion line in the $M\pi \to M\pi$ amplitude diagram, leading to 
\begin{eqnarray}\label{Eq:Pi}
 \Pi_M(p^0,\vec p;T) = \int\!\frac{\mathrm d^3q}{(2\pi)^3} \int\!\mathrm d\Omega
 \frac{ f(\Omega,T)-f(\omega_\pi,T)}{(p^0)^2 - (\omega_\pi-\Omega)^2 + {\rm i}\varepsilon}  
 \left(-\frac1\pi\right) \mathrm{Im} T_{M \pi}(\Omega,\vec p+\vec q;T) \ .
\end{eqnarray}

\section{Results}

\begin{figure*}[ht]
\centering
  \includegraphics[width=0.85\linewidth]{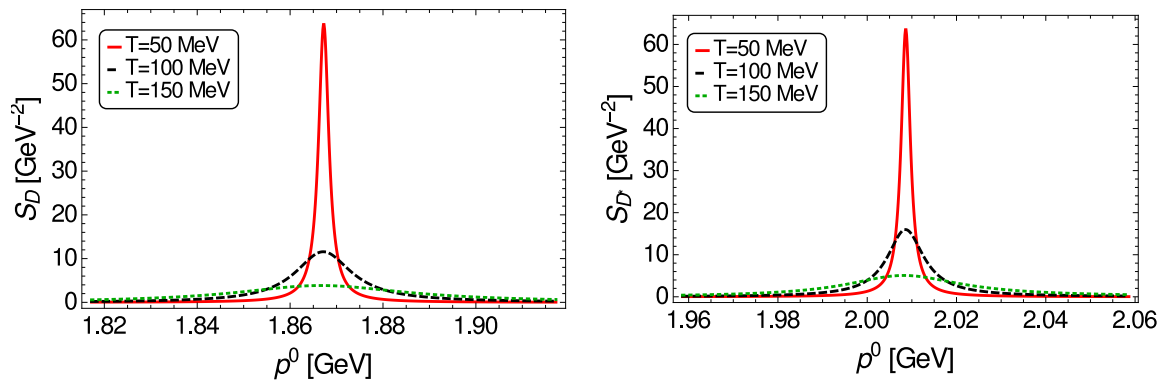}
\caption{Meson spectral function as a function of the energy $p^0$ at temperatures $T = 50$, 100, 150 MeV and $\vec{p} = 0$. a): $D$, (b): $D^*$ .} 
\label{Fig:SpecDDstar}
\end{figure*}

The self-consistent determination of the meson self-energies has been done in Ref.~\cite{Cleven:2017fun} and we show here the results for the most relevant mesons, $D$ and $D^*$. Their zero momentum spectral functions are shown in Fig.~\ref{Fig:SpecDDstar} as functions of energy $p^0$ for various temperatures. As already commented in Ref.~\cite{Cleven:2017fun}, the shift of the spectral function peak, related to the real part of the self-energy,  is negligible, while the width, connected to the imaginary-part, becomes substantially larger as temperature increases. 

\begin{figure*}[ht]
\centering
  \includegraphics[width=0.85\linewidth]{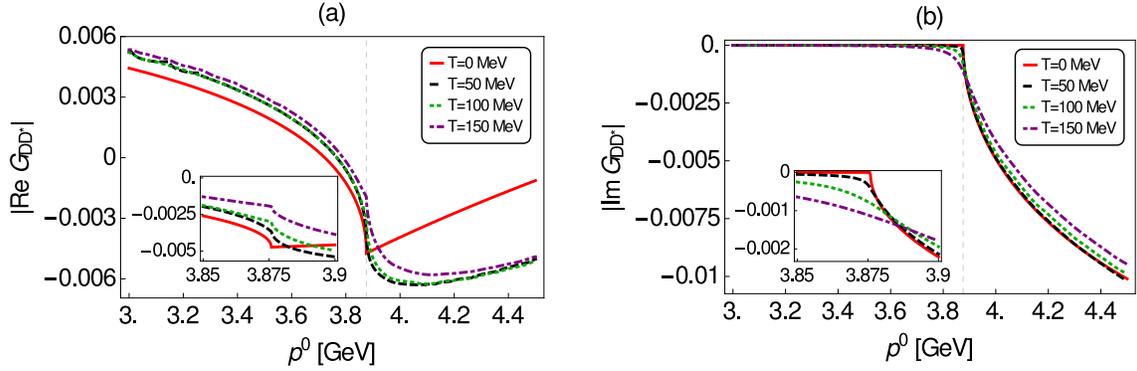}
\caption{Real part of the $DD^*$ loop (left panel), imaginary part of the $DD^*$ loop (right panel).  
All quantities are shown at temperatures 0, 50, 100, 150~MeV. The dashed gray line represents the $DD^*$ threshold} .
\label{Fig:DDstar}
\end{figure*}

%
In the following we discuss the effect the finite temperature spectral functions of the charmed mesons have in the loop function and, consequently, in the amplitude $T_{D D^*}$ which signals the generation of the $\x$.

The left and right panels of Fig.~\ref{Fig:DDstar} show the real and imaginary parts of the $DD^*$ loop, respectively, as functions of the total energy $P^0$ for a total momentum $\vec{P}=0$ and various temperatures.
Below threshold we find that the finite temperature real part retains its shape but is slightly shifted compared to the vacuum case, increasingly so with increasing temperature. 
Above threshold the effect of the hot pion bath becomes more pronounced.
Instead of the sharp rise after the kink at threshold observed in vacuum, we see a smoother behaviour and a slower rise at finite temperature, resulting in differences of about a factor two in size far above threshold. However, this is beyond the energy region of interest with regard to the $\x$.
For the imaginary part of the loop we see a similar behaviour. The sharp opening of the unitarity cut at zero temperature is transformed into a smoother curve at finite temperature. It is interesting to note that there is a substantial strength in a range of a few tens of MeV below threshold, right where the vacuum imaginary part vanishes. 
This is quite significant since it means that the energy region where the pole of the $\x$ is located is affected strongly by the surrounding pions. 
Thus the loosely bound $\x$ is an excellent candidate to study the effect of the hot medium compared to more tightly bound states. 

The noticeable effect of temperature on the $D D^*$ loop around threshold is due to an important modification of the $D\pi$ and $D^*\pi$ interactions in a hot medium composed essentially of pions. It is precisely the self-energy of the $D$ and $D^*$ mesons what makes  their corresponding spectral functions acquire a substantial width, affecting the value of the $D D^*$ loop especially around threshold.
We note that the null contributions of the diagonal $D_s\pi$ and $D^*_s\pi$ potentials \cite{Gamermann:2006nm,Gamermann:2007fi} make these unitarized interactions substantially weaker than their non-strange $D\pi$ and $D^*\pi$ counterparts, which would be reflected into narrower $D_s$ and $D^*_s$ spectral functions. This, together with the fact that the $ D_s D_s^*$ channel lies about 300 MeV above the relevant region of interest for the $\x$ meson, justifies evaluating the $G_{D_s D^*_s} $ loop with delta-type distributions for the $D_s$ and $D^*_s$ spectral functions.

\begin{figure*}[ht]
\centering
  \includegraphics[width=0.5\linewidth]{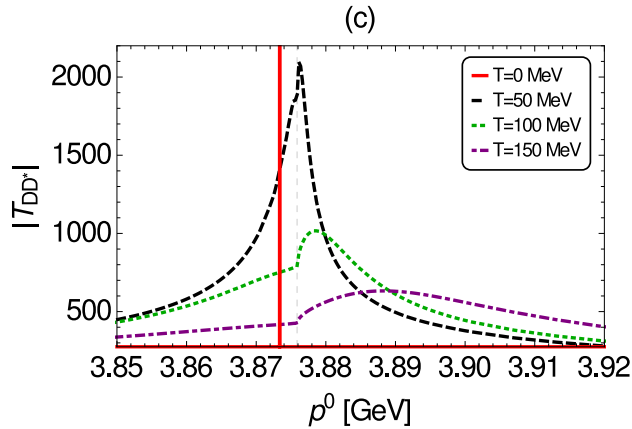}
\caption{Absolute value of the unitarized amplitude for $DD^*$ scattering at temperatures 0, 50, 100, 150~MeV. The dashed gray line represents the $DD^*$ threshold, the vertical line represents the stable $\x$ at $T=0~\mev$ .
} 
\label{Fig:XTemp}
\end{figure*}

The solution of Eq.~(\ref{Eq:T}) with hot medium modified loops gives rise to unitarized amplitudes from which the properties of the $\x$ can be extracted. 
In Fig.~\ref{Fig:XTemp} we show the absolute value of the $D D^*$ scattering amplitude for various temperatures. 
These results illustrate the impact that the modifications of the loop function discussed in the previous paragraphs have on the dynamical generation of the $\x$. 
The most important observation is that the peak associated to the $\x$ becomes significantly wider with increasing temperature. This is due to the combination of two effects. Firstly, the peak broadens because of the appearance of a finite imaginary part of the amplitude at the pole position when temperature effects are included. Secondly, the repulsive shift in the real part of the loop below and around threshold, as clearly seen in the inset of Fig.~\ref{Fig:DDstar}(a), implies that the resonance is generated at higher energies, moving the peak position from slightly below threshold in vacuum to some MeV above it at finite temperature. This shift increases the width of the $\x$ tremendously since it can decay into most of the $D$ and $D^*$ meson spectral strength.  

From the line shape of the amplitude we derive the $\x$ width, going from being 
stable at zero temperature to values $\sim$10, $\sim$30 and $\sim$60 MeV at temperatures 50, 100 and 150 MeV, respectively.
Thus, at typical kinetic freeze out temperatures for RHIC and LHC, the $\x$ can not be considered as a loosely bound (deutron-like) $D D^*$ bound state.
For example, in the calculations of the ExHIC collaboration \cite{Cho:2017dcy} the freeze out temperature was established between 115 and 119 MeV and out results indicate that the  $\x$  would be converted into a resonance with a width of about 40 MeV. In fact, during the whole hadronic phase of the reaction the width of $\x$ would slowly be changing from about 65 MeV at the hadronization temperature (162~MeV at the highest RHIC energy and 156~MeV at LHC \cite{Cho:2017dcy}) to 40 MeV at freeze out. 
Our expectation is that such a displacement of the $\x$ peak above the threshold together with the related broadening of its shape should translate, according to the hadron coalescence model, into a lower estimation of production yields for a molecular-like $\x$ state in energetic relativistic heavy-ion collisions.

\section{Summary and Conclusions}

In this work we have explored the properties of the $\x$ in a hot pionic medium assuming this resonance to be a molecular state generated by the interaction of $D \bar D^* + c.c.$ pairs and associated coupled channels. The model employed considers the broadening of the $D$ and $D^*$ mesons in the pionic medium as a  consequence of their self-energies, developed from the corresponding  $D\pi$ and $D^*\pi$ temperature dependent amplitudes. Once the properties of the $D$ and $D^*$ mesons in a hot pionic medium are incorporated in the loop functions of the Bether-Salpeter equation, the properties of the generated $\x$ can be inferred from the resulting $D D^*$ amplitude.

We find a substantial broadening of the $\x$ of a few tens of MeV when temperature effects are considered. The behaviour of the $\x$ presented here is a unique feature of the molecular interpretation and is due to the relatively strong interactions of the charmed mesons with the pion bath. A tetraquark-type state would barely change its behaviour under the same circumstances. 

It has been argued that if the $\x$ is a molecular state it will be produced by hadron coalescence at the end of the hadronic phase with yields at least one order of magnitude higher than the surviving abundance, in the hadronic phase, of tetraquark-type states produced at the mixed phase by quark coalescence \cite{Cho:2010db,Cho:2011ew,Cho:2017dcy,Abreu:2016qci}. These estimations would need to be revisited in view of the findings of the present letter, as the temperature dependent widening of the $\x$ established here will influence the predictions of its production yield in the molecular scenario. 
We hope that the consideration of our results would help in finding a more realistic prediction of the heavy-ion collision observables that have been argued to be additional indicators for establishing the nature of the $\x$.

\section*{Acknowledgments}
This work is partly supported by the Spanish Ministerio de Economia y Competitividad (MINECO) under the project MDM-2014-0369 of ICCUB (Unidad de Excelencia 'Mar\'\i a de Maeztu'), 
and, with additional European FEDER funds, under the contracts FIS2014-54762-P and
FIS2017-87534-P.



\begin{thebibliography}{99}


\bibitem{Choi:2003ue}
  S.~K.~Choi {\it et al.} [Belle Collaboration],
  Phys.\ Rev.\ Lett.\  {\bf 91}, 262001 (2003)
  doi:10.1103/PhysRevLett.91.262001
  [hep-ex/0309032].

\bibitem{Acosta:2003zx}
  D.~Acosta {\it et al.} [CDF Collaboration],
  Phys.\ Rev.\ Lett.\  {\bf 93} (2004) 072001
  doi:10.1103/PhysRevLett.93.072001
  [hep-ex/0312021].
 
\bibitem{Abazov:2004kp}
  V.~M.~Abazov {\it et al.} [D0 Collaboration],
  Phys.\ Rev.\ Lett.\  {\bf 93} (2004) 162002
  doi:10.1103/PhysRevLett.93.162002
  [hep-ex/0405004].
  
\bibitem{Aubert:2004ns}
  B.~Aubert {\it et al.} [BaBar Collaboration],
  Phys.\ Rev.\ D {\bf 71} (2005) 071103
  doi:10.1103/PhysRevD.71.071103
  [hep-ex/0406022].
  
\bibitem{Aaij:2011sn}
  R.~Aaij {\it et al.} [LHCb Collaboration],
  Eur.\ Phys.\ J.\ C {\bf 72} (2012) 1972
  doi:10.1140/epjc/s10052-012-1972-7
  [arXiv:1112.5310 [hep-ex]].
  
  
\bibitem{mol1} 
  E.~S.~Swanson,
  Phys.\ Rept.\  {\bf 429}, 243 (2006)
  doi:10.1016/j.physrep.2006.04.003
  [hep-ph/0601110].

\bibitem{mol2} 
  D.~Gamermann and E.~Oset,
  Phys.\ Rev.\ D {\bf 80}, 014003 (2009)
  doi:10.1103/PhysRevD.80.014003
  [arXiv:0905.0402 [hep-ph]].
 
 
\bibitem{tetra1} 
  L.~Maiani, F.~Piccinini, A.~D.~Polosa and V.~Riquer,
  Phys.\ Rev.\ D {\bf 71}, 014028 (2005)
  doi:10.1103/PhysRevD.71.014028
  [hep-ph/0412098].
  
\bibitem{tetra2} 
  R.~D.~Matheus, S.~Narison, M.~Nielsen and J.~M.~Richard,
  Phys.\ Rev.\ D {\bf 75}, 014005 (2007)
  doi:10.1103/PhysRevD.75.014005
  [hep-ph/0608297].
  
  
\bibitem{combi_2_4} 
  R.~D.~Matheus, F.~S.~Navarra, M.~Nielsen and C.~M.~Zanetti,
  Phys.\ Rev.\ D {\bf 80}, 056002 (2009)
  doi:10.1103/PhysRevD.80.056002
  [arXiv:0907.2683 [hep-ph]].
 
  
\bibitem{Guo:2017jvc}
  F.~K.~Guo, C.~Hanhart, U.~G.~Meißner, Q.~Wang, Q.~Zhao and B.~S.~Zou,
  arXiv:1705.00141 [hep-ph].
  
\bibitem{Brambilla:2010cs}
  N.~Brambilla, S.~Eidelman, B.~K.~Heltsley, R.~Vogt, G.~T.~Bodwin, E.~Eichten, A.~D.~Frawley and A.~B.~Meyer {\it et al.},
  Eur.\ Phys.\ J.\ C {\bf 71}, 1534 (2011)
  [arXiv:1010.5827 [hep-ph]].

  
\bibitem{Esposito:2014rxa}
  A.~Esposito, A.~L.~Guerrieri, F.~Piccinini, A.~Pilloni and A.~D.~Polosa,
  Int.\ J.\ Mod.\ Phys.\ A {\bf 30} (2015) 1530002
  doi:10.1142/S0217751X15300021
  [arXiv:1411.5997 [hep-ph]].


 
\bibitem{Bignamini:2009sk} 
  C.~Bignamini, B.~Grinstein, F.~Piccinini, A.~D.~Polosa and C.~Sabelli,
  Phys.\ Rev.\ Lett.\  {\bf 103}, 162001 (2009)
 

\bibitem{Carvalho:2015nqf}
  F.~Carvalho, E.~R.~Cazaroto, V.~P.~Goncalves and F.~S.~Navarra,
  Phys.\ Rev.\ D {\bf 93}, no. 3, 034004 (2016).


   


  
\bibitem{Cho:2010db}
  S.~Cho {\it et al.} [ExHIC Collaboration],
  Phys.\ Rev.\ Lett.\  {\bf 106} (2011) 212001
  doi:10.1103/PhysRevLett.106.212001
  [arXiv:1011.0852 [nucl-th]].

\bibitem{Cho:2011ew}
  S.~Cho {\it et al.} [ExHIC Collaboration],
  Phys.\ Rev.\ C {\bf 84} (2011) 064910
  doi:10.1103/PhysRevC.84.064910
  [arXiv:1107.1302 [nucl-th]].
  
\bibitem{Cho:2017dcy}
  S.~Cho {\it et al.} [ExHIC Collaboration],
  Prog.\ Part.\ Nucl.\ Phys.\  {\bf 95} (2017) 279
  doi:10.1016/j.ppnp.2017.02.002
  [arXiv:1702.00486 [nucl-th]].



\bibitem{Cho:2013rpa}
  S.~Cho and S.~H.~Lee,
  Phys.\ Rev.\ C {\bf 88} (2013) 054901
  doi:10.1103/PhysRevC.88.054901
  [arXiv:1302.6381 [nucl-th]].


  
\bibitem{Abreu:2016qci}
  L.~M.~Abreu, K.~P.~Khemchandani, A.~Martinez Torres, F.~S.~Navarra and M.~Nielsen,
  Phys.\ Lett.\ B {\bf 761} (2016) 303
  doi:10.1016/j.physletb.2016.08.050
  [arXiv:1604.07716 [hep-ph]].
  
\bibitem{Torres:2014fxa}
  A.~Martinez Torres, K.~P.~Khemchandani, F.~S.~Navarra, M.~Nielsen and L.~M.~Abreu,
  Phys.\ Rev.\ D {\bf 90} (2014) no.11,  114023
   Erratum: [Phys.\ Rev.\ D {\bf 93} (2016) no.5,  059902]
  doi:10.1103/PhysRevD.93.059902, 10.1103/PhysRevD.90.114023
  [arXiv:1405.7583 [hep-ph]].
  

\bibitem{Cleven:2017fun}
  M.~Cleven, V.~K.~Magas and A.~Ramos,
  Phys.\ Rev.\ C {\bf 96} (2017) no.4,  045201
  doi:10.1103/PhysRevC.96.045201
  [arXiv:1707.05728 [hep-ph]].





\bibitem{Gamermann:2006nm}
  D.~Gamermann, E.~Oset, D.~Strottman and M.~J.~Vicente Vacas,
  Phys.\ Rev.\ D {\bf 76} (2007) 074016
  doi:10.1103/PhysRevD.76.074016
  [hep-ph/0612179].


\bibitem{Gamermann:2007fi}
  D.~Gamermann and E.~Oset,
  Eur.\ Phys.\ J.\ A {\bf 33} (2007) 119
  doi:10.1140/epja/i2007-10435-1
  [arXiv:0704.2314 [hep-ph]].
  





\bibitem{galekapustabook}  J.I. Kapusta and C.Gale, ``Finite temperature field theory. Principles and Applications''. Cambridge University Press 2006.

\bibitem{lebellac} M.Le Bellac, {\it Thermal Field Theory} (Cambridge
University Press, 1996).
  
\end{thebibliography}
\end{document}